\documentstyle{article}
\input phyzzx.fonts
\def\and{{\it\&}}

\def\to{\rightarrow}
\def\gesim{\,{\raise-3pt\hbox{$\sim$}}\!\!\!\!\!{\raise2pt\hbox{$>$}}\,}
\def\lesim{\,{\raise-3pt\hbox{$\sim$}}\!\!\!\!\!{\raise2pt\hbox{$<$}}\,}
\def\boldoverdot{\,{\raise6pt\hbox{\bf.}\!\!\!\!\>}}

\def\ibid{{\it ibid.}\ }

\def\lcal{{\cal L}}

\def\ocal{{\cal O}}

\def\vcal{{\cal V}}
\def\wcal{{\cal W}}

\def\diag{\hbox{\diag}}

\def\gev{\hbox{GeV}}

\def\Dslash#1{\slash\!\!\!\!#1}
\def\inbox#1{\vbox{\hrule\hbox{\vrule\kern5pt
     \vbox{\kern5pt#1\kern5pt}\kern5pt\vrule}\hrule}}
\def\sqr#1#2{{\vcenter{\hrule height.#2pt
      \hbox{\vrule width.#2pt height#1pt \kern#1pt
         \vrule width.#2pt}
      \hrule height.#2pt}}}

\def\today{\ifcase\month\or
  January\or February\or March\or April\or May\or June\or
  July\or August\or September\or October\or November\or December\fi
  \space\number\day, \number\year}
\def\pmb#1{\setbox0=\hbox{#1}%
  \kern-.025em\copy0\kern-\wd0
  \kern.05em\copy0\kern-\wd0
  \kern-.025em\raise.0433em\box0 }
\def\antes{}
\def\despues{.}
\def\sumprime_#1{\setbox0=\hbox{$\scriptstyle{#1}$}
  \setbox2=\hbox{$\displaystyle{\sum}$}
  \setbox4=\hbox{${}'\mathsurround=0pt$}
  \dimen0=.5\wd0 \advance\dimen0 by-.5\wd2
  \ifdim\dimen0>0pt
  \ifdim\dimen0>\wd4 \kern\wd4 \else\kern\dimen0\fi\fi
\mathop{{\sum}'}_{\kern-\wd4 #1}}
\def\sumbiprime_#1{\setbox0=\hbox{$\scriptstyle{#1}$}
  \setbox2=\hbox{$\displaystyle{\sum}$}
  \setbox4=\hbox{${}'\mathsurround=0pt$}
  \dimen0=.5\wd0 \advance\dimen0 by-.5\wd2
  \ifdim\dimen0>0pt
  \ifdim\dimen0>\wd4 \kern\wd4 \else\kern\dimen0\fi\fi
\mathop{{\sum}''}_{\kern-\wd4 #1}}
\def\sumtriprime_#1{\setbox0=\hbox{$\scriptstyle{#1}$}
  \setbox2=\hbox{$\displaystyle{\sum}$}
  \setbox4=\hbox{${}'\mathsurround=0pt$}
  \dimen0=.5\wd0 \advance\dimen0 by-.5\wd2
  \ifdim\dimen0>0pt
  \ifdim\dimen0>\wd4 \kern\wd4 \else\kern\dimen0\fi\fi
\mathop{{\sum}'''}_{\kern-\wd4 #1}}
\newcount\chapnum

\def\chap#1{\clearsect\clearprob
\global\advance\chapnum by 1 \par\vskip .5 in\par%
\centerline{{\bigboldiii\antes\the\chapnum\despues\ #1}}}
\newcount\sectnum
\def\clearsect{\sectnum=0}
\def\sect#1{\clearprob\global\advance\sectnum by 1 \par\vskip .25 in\par%
\noindent{\bigboldii\the\chapnum.\the\sectnum:\ #1}\nobreak}
\newcount\yesnonum

\def\verify{\global\advance\yesnonum by 1{\bigboldi (VERIFY!!)}}
\def\tocheck{\par\vskip 1 in{\bigboldv TO VERIFY: \the\yesnonum\ ITEMS.}}
\newcount\notenum

\def\note#1{\global\advance\notenum by 1{ \bf $<<$ #1 $>>$ } }
\def\noteout{\par\vskip 1 in{\bigboldiv NOTES: \the\notenum.}}
\newcount\borrownum

\def\borrow{\global\advance\borrownum by 1{\bigboldi BORROWED BY:\ }}
\def\borrowed{\par\vskip 0.5 in{\bigboldii BOOKS OUT:\ \the\borrownum.}}
\newcount\refnum

\def\ref#1{\global\advance\refnum by 1\item{\the\refnum.\ }#1}
\def\stariref#1{\global\advance\refnum by 1\item{%
               {\bigboldiv *}\the\refnum.\ }#1}
\def\stariiref#1{\global\advance\refnum by 1\item{%
               {\bigboldiv **}\the\refnum.\ }#1}
\def\stariiiref#1{\global\advance\refnum by 1\item{
               {\bigboldiv ***}\the\refnum.\ }#1}
\newcount\probnum
\def\clearprob{\probnum=0}
\def\prob{\global\advance\probnum by 1 {\medskip $\triangleright$\
\undertext{{\sl Problem}}\ \the\chapnum.\the\sectnum.\the\probnum.\ }} 
\newcount\probchapnum

\def\probchap{\global\advance\probchapnum by 1 {\medskip $\triangleright$\
\undertext{{\sl Problem}}\ \the\chapnum.\the\probchapnum.\ }} 
\def\undertext#1{$\underline{\smash{\hbox{#1}}}$}

\newskip\humongous \humongous=0pt plus 1000pt minus 1000pt
\def\caja{\mathsurround=0pt}
\def\eqalign#1{\,\vcenter{\openup1\jot \caja
	\ialign{\strut \hfil$\displaystyle{##}$&$
	\displaystyle{{}##}$\hfil\crcr#1\crcr}}\,}
\newif\ifdtup

\def\oldreffmt#1{\rlap{[#1]} \hbox to 2\parindent{}}

\def\figfmt#1{\rlap{Figure {#1}} \hbox to 1in{}}

\def\bra#1{\left\langle #1\right|}
\def\ket#1{\left| #1\right\rangle}

\def\beq{\begin{equation}}
\def\eeq{\end{equation}}

\relax

\def\sutwoone{SU_2^L\otimes U_1^Y}
\def\lqed{\lcal_{QED}}
\def\leff{\lcal_{eff}}
\def\lnr{\lcal_{NR}}
\def\lsm{\lcal_{SM}}
\def\lhiggs{\lcal_H}
\def\nextline{\unskip\nobreak\hskip\parfillskip\break}

\def\mn{{\mu\nu}}

\def\lsim{\mathop <\limits_{\raise2.3pt\hbox{$\sim$}}}
\def\gsim{\mathop >\limits_{\raise2.3pt\hbox{$\sim$}}}

\font\tenrm=cmr10
\font\tenit=cmti10
\font\elevenbf=cmbx10 scaled\magstep 1
\font\elevenrm=cmr10 scaled\magstep 1
\font\elevenit=cmti10 scaled\magstep 1

\font\ninerm=cmr9
\font\nineit=cmti9

\renewenvironment{thebibliography}[1]
 { \elevenrm
   \begin{list}{\arabic{enumi}.}
    {\usecounter{enumi} \setlength{\parsep}{0pt}
     \setlength{\itemsep}{3pt} \settowidth{\labelwidth}{#1.}
     \sloppy
    }}{\end{list}}

\begin{document}
\begin{flushright}
UM--TH--93--17
hep-ph/9308331
\end{flushright}
\begin{center}
\vglue 0.6cm
{{\elevenbf THEORY OF GAUGE BOSON STRUCTURE\\}
\vglue 5pt
{\tenrm MARTIN B. EINHORN\footnote{\ninerm\baselineskip=11pt Based in part on
work done in collaboration with J. Wudka and C. Arzt.} \\}
\baselineskip=13pt
{\tenit Randall Laboratory of Physics, University of Michigan\\}
\baselineskip=12pt
{\tenit Ann~Arbor, MI  48109-1120, USA\\}
}
\end{center}
\vglue 0.6cm

\def\twoone{SU_2^L\otimes U_1^Y}

{\elevenbf\noindent 1. Introduction}
\vglue 0.2cm
\baselineskip=14pt
\elevenrm

I have been asked to review the theory of vector boson structure.  Recent
discussion about the possibility of composite vectors and the meaning of
structure has been rather confusing, but I believe a consensus is beginning to
emerge.  Because the following talk by Dr. Miyamoto will review in some detail
the phenomenological possibilities, I will concentrate on the theoretical
underpinnings.  In the context of the Next Linear Collider, I would say the
main goal is to understand the extent to which studies of vector boson
couplings can provide insights into new, unknown physics beyond the Standard
Model, given the  sensitivities that are likely to be achievable.  The outline
of my talk is as follows:  In the next section, in order to introduce my
nomenclature and to highlight the issues in a simpler context, I begin by
considering the analogous situation for QED.  In Section~3, I will generalize
to the present-day situation.  In Section~4, I review the now-standard
parameterization of deviations of the triple-vector-boson couplings from the
SM.  In Section~5, I will expand on the case of the strongly interacting Higgs
sector.  In Section~6, I discuss the orders of magnitude to be expected on
general theoretical grounds.  In Section~7, I will try to indicate why gauge
invariance is such a critical issue and why it would be so difficult to
understand the present successes of the SM without
it.\footnote{\ninerm\baselineskip=11pt Some of the observations in this talk
appeared originally in \cite{hiro}.}

\vglue 0.6cm
{\elevenbf\noindent 2. Before the Standard Model}
\vglue 0.4cm
\setcounter{footnote}{1}

One topic that has been debated is whether one must demand gauge invariance
for the interactions of the vector bosons.  There are many dimensions of this
issue, and it is both pedagogically helpful and theoretically illuminating to
begin with a somewhat simpler situation than we face today.  Long before the
Standard Model (SM) had been formulated, there was Quantum Electrodynamics
(QED), which may be summarized by the
Lagrangian:\footnote{\ninerm\baselineskip=11pt Generally, the gauge-fixing
term $\lcal_{g.f.}$ is chosen so that the Faddeev-Popov ghosts in
$\lcal_{F-P}$ decouple, but with a view toward the non-Abelian case, we've
included that term here.}
\beq
\eqalign{\lqed\equiv &\>\lcal_g+\lcal_m,\cr
\lcal_g\equiv &-{1\over4}F_\mn^2+\lcal_{g.f.}+\lcal_{F-P},\qquad
\lcal_m\equiv \sum_n\overline{\psi_n}(i\Dslash{D}-m_n)\psi_n\cr
F_\mn\equiv &\>\partial_\mu A_\nu-\partial_\nu A_\mu,\qquad
D_\mu\psi_n\equiv\> (\partial_\mu+ie_n A_\mu )\psi_n,\cr}
\eeq
where $A_\mu$ denotes the photon field and the $\psi_n$ denote the various
matter fields.  In ancient times, they were the electron and muon, nucleon and
pion fields.  (One may include the quarks and their interactions, now
known to be described by QCD.)  The action associated with this Lagrangian is
gauge invariant under the $U_1^Q$ symmetry of electromagnetism.  I want to
recall two important things that are accomplished by this gauge symmetry:

{(1)}~~To describe a massless photon, one wants only two physical degrees
of freedom of the four-components of $A_\mu.$  This is accomplished, first,
because the time-derivative of $A_0$ does not occur, except through the gauge
fixing term, so that $A_0$ is not a physical dynamical variable.  This is
important because this field would create
ghosts,\footnote{\ninerm\baselineskip=11pt Ghosts $\equiv$ states of negative
norm.}  and gauge invariance insures that this feature is preserved by
radiative corrections.  Secondly, because of the gauge symmetry, of the
remaining 3 dynamical space-components $A_i$, only two are physical, as can be
seen by choosing a gauge (such as the Lorentz or Coulomb gauges) which imposes
a constraint among them.  I remind you of these well-known facts because
tampering with gauge invariance by adding terms that explicitly break the
gauge symmetry can be very dangerous and generally will imperil these
important features.

{(2)}~~A second important aspect of gauge invariance is universality.  We
are familiar with the fact that Lagrangian parameters are not directly
observable, but gauge invariance insures that the ratio of physical charges
equal the ratios of Lagrangian charges.\footnote{In textbooks, this usually
is associated with the equality of the charge and matter wave-function
renormalization constants $Z_1=Z_2.$}  This is why, if the electron and
muon have the same charge in the Lagrangian, they have the same observed
charges. This feature is preserved under the addition of strong
interactions and  protons (or quarks) are included with charge proportional
to the electron charge.  While it was not understood why the proton's
charge was equal and opposite to the electron's charge, at least it was
understood that if it were true for the Lagrangian parameters, it would be
true for the physical charges as well.  This relationship is extremely
important, being ultimately responsible for the neutrality of atoms.
Again, explicit breaking of gauge invariance would render it a great
mystery and challenge for theory to explain these fundamental facts of
nature.
\vglue 0.2cm
There is a third point that is often times also cited in this connection.
Frequently, it is said that gauge invariance forbids the occurrence of a
photon mass term $m_\gamma^2 A_\mu^2$, and keeps the photon massless.  In this
form it is true, but gauge invariance does not really prevent the photon from
getting a mass, as St\"uckelberg pointed out in 1938.\cite{stuck}  One may add
to the theory a scalar field $\chi$ and a term to the Lagrangian of the form
\beq
\lcal_m\equiv\>(mA_\mu-\partial_\mu\chi)^2.
\eeq
The $U_1^Q$ gauge symmetry is maintained by extending it to $\chi$:
\beq
\eqalign{A_\mu&\to A_\mu+\partial_\mu\Theta,\>\> {\rm vector\>gauge\>field}\cr
\chi&\to \chi+m\Theta,\>\> {\rm scalar\>gauge\>field},\cr}
\eeq
The field $\chi$ is sometimes referred to as the St\"uckelberg field, but we
might call it a ``scalar" gauge field by analogy with $A_\mu.$  While $\chi$
adds one degree of freedom to the theory, if this is the only place it enters,
it may be regarded as a gauge artifact.  In the ``unitary gauge," ($\chi=0$)
it disappears, and this term takes on the appearance of a mass term for the
vector field.  Indeed, this is the correct interpretation, showing that one
can have a gauge-invariant, massive photon.\footnote{\ninerm\baselineskip=11pt
This theory retains renormalizability.  In the SM, one could do the same thing
for the hypercharge field.  As we shall see below, the similar trick in the
non-Abelian case results in a non-renormalizable theory.} I think it is fair
to say that we still don't really understand why the photon is massless,
although the fact that it is may be a hint at grand unification.

Whereas the leptons were pointlike, nucleons had structure, indicated by the
presence of form factors:
\beq
\eqalign{\bra{p}J_\mu^{em}\ket{p}=&e\overline{u(p')}[\gamma_\mu
F_1(q^2)+{{i\sigma_\mn q^\nu}\over{2M}}F_2(q^2)]u(p),\cr}
\eeq
where $q\equiv p-p'$.  The proton and neutron were found to have large
magnetic moments:
\beq
F^p_2(0)=1.8,\>\>F^n_2(0)=-1.9.
\eeq
This was interpreted as the effects of the strong interactions, but, as the
theme of this talk is the nature of structure, one might ask why one simply
did not add Pauli-type interactions to the original Lagrangian
\beq
{{\alpha_{\psi F}}\over\Lambda} \overline{\psi}\sigma_\mn\psi F^\mn
\eeq
At the phenomenological level, the form factors for nucleons were soon found
to be rapidly falling functions of $q^2$, but, at a more theoretical level,
such an interaction term was objectionable because it was not renormalizable.
Although it has not yet been experimentally verified, we now believe
that, even after including a correct theory of strong interactions to account
for the hadronic form factors, such local Pauli-interactions
should be included in QED, even for leptons!  They arise from weak vector
boson effects, such as depicted in Fig.~1, and appear to be local interactions
at scales far below the weak vector boson masses.
\vskip0.2in
\input epsf

\setbox1=\vbox {\hsize=1.5truein
\epsfxsize=\hsize
\epsffile{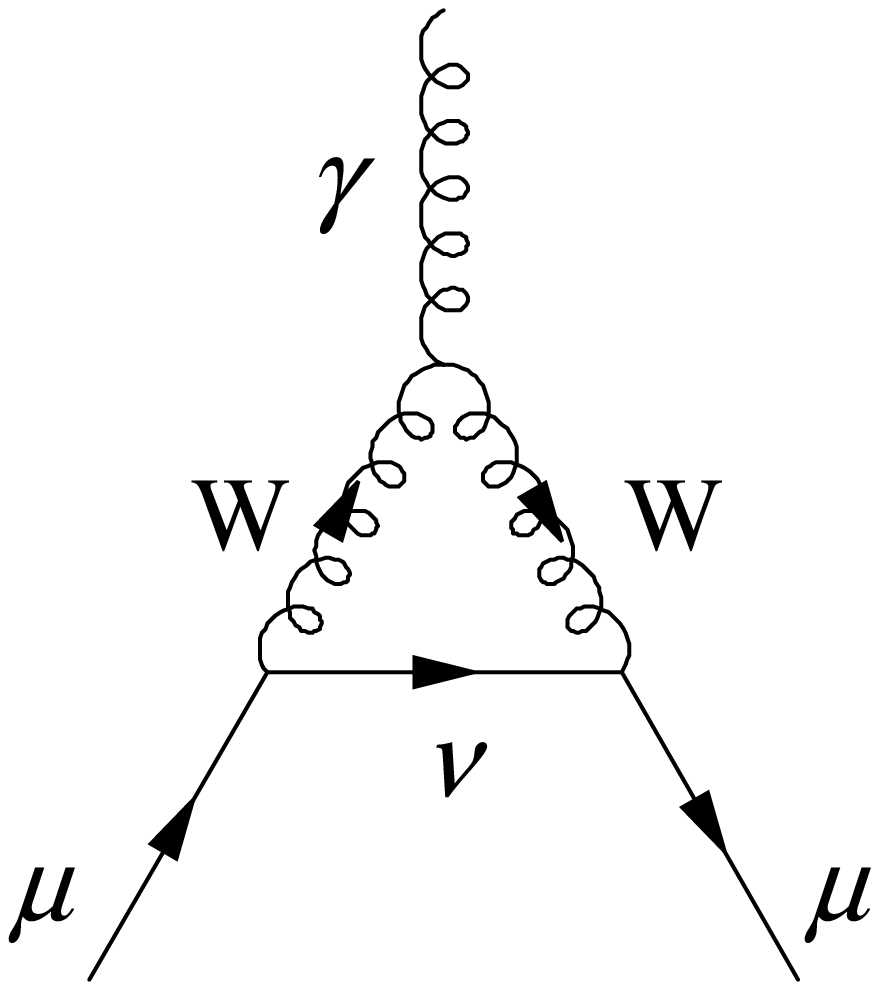}}

\setbox2=\vbox {\hsize=1.5truein
\epsfxsize=\hsize
\epsfysize=\ht1
\epsffile{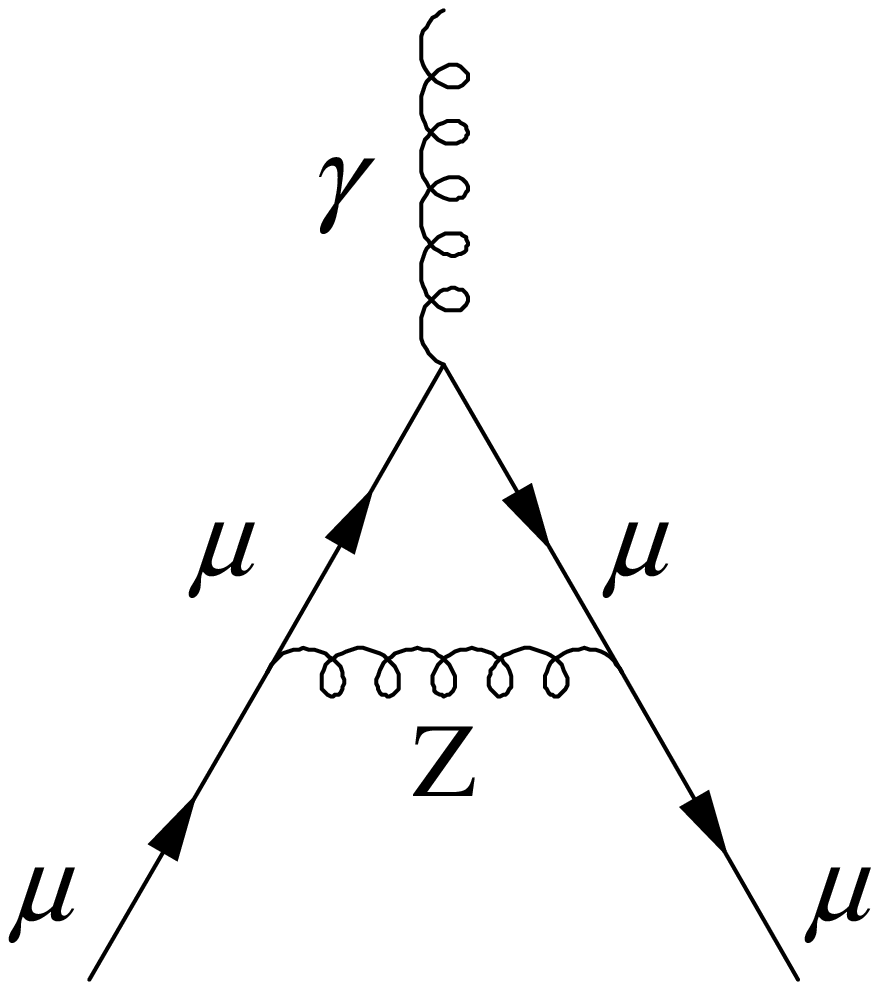}}

\vbox {\hbox to \hsize { \hfill \box1 \hfill \box2 \hfill}}


\centerline{\tenrm Weak boson contributions to the anomalous magnetic
moment.}
\nobreak\centerline{\tenrm Fig.~1}
\vskip.2in
In fact, this illustrates what must be expected in any theory short of the
ultimate theory.  ``Nonrenormalizable," higher-dimensional interactions are
generally present resulting from new physics at a higher energy scale and
above which the present form of the local field theory breaks down, usually
because new physical degrees of freedom come into play.   For example, in
Eq.~(6), the scale $\Lambda$ may be identified with the vector boson
masses.  If one regards $\Lambda$ as a physical cutoff where the present
theory breaks down, then these higher-dimensional operators do not spoil
renormalizability in the following sense:  For consistency, one must imagine
that the Lagrangian includes all such higher dimensional operators allowed by
the symmetries. Such a representation is called an {\elevenit effective field
theory,}  and I am suggesting that any realistic theory (short of the
ultimate) will be of this type.  Then, although the inclusion of such
``nonrenormalizable" vertices in loop corrections lead to divergences in
arbitrary Green's functions, all such divergences may be absorbed in
relationships between Lagrangian parameters and observables.

This does imply, however, that, unlike a renormalizable theory, there are
really an infinity of coupling constants to be fit to experiment, so you may
well wonder how it is possible to predict anything.  For example, given that
there is a fundamental interaction of the Pauli type, how is it possible for
QED to predict correctly the value of $g-2$ for the electron and muon with
such high precision?  The answer is that it depends on the size of the
nonrenormalizable vertex coming from the heavy sector as compared with the
size of the radiative corrections coming from the light theory.   The most
important corrections to QED predictions actually come from hadronic effects,
but, for pedagogical purposes, let me suppose that these can be precisely
determined, as, in fact, they can be.\footnote{\ninerm\baselineskip=11pt This
has been reviewed in Ref.~\cite{kinoshita}, although there remain certain
uncertainties due to the so-called hadronic light-by-light scattering
corrections that may be as large as the weak corrections.}   It is
illustrative for later comparison with corrections to the SM to estimate how
large are the corrections coming from the weak loops in Fig.~1.   If one
expands the weak corrections in terms of the vector boson masses, there are
logarithmic dependences that can be absorbed in the counterterm for the muon
charge, but there are finite corrections to the magnetic moment Eq.~(6) that
may be estimated as follows: For example, the W-boson contribution is the
product of three factors:
\beq
{{\alpha_{\mu F}}\over\Lambda}=\Bigl({e\over{M_W}}\Bigr)
\Bigl({{g^2}\over{16\pi^2}}\Bigr)
\Bigl({{m_\mu}\over{M_W}}\Bigr).
\eeq
Each factor has significance:  $e/{M_W}$ represents the strength of photon
coupling to $A_\mu$ over scale of new physics; ${{g^2}/{16\pi^2}}$
represents a new interaction strength, $g$, times a loop-factor of
$1/(16\pi^2)$;  a further suppression factor of ${{m_\mu}/{M_W}}$ due to
the  chiral structure of electroweak interactions.  This last suppression
because of a global symmetry is fortunate for tests of QED, since it means
that Pauli-term acts more like a dimension-six rather than a dimension-five
operator.\footnote{\ninerm\baselineskip=11pt As small as this is, an
experiment being constructed at BNL hopes to measure $g_\mu-2$ with a
design sensitivity about 5 times smaller than the size of the predicted
weak corrections, so this discussion is quite relevant.}  In the common
parlance, this contribution is a correction to ${{(g_\mu-2)}/{2m_\mu}}$,
even though the nominal scale of the contribution is not set by the muon
mass but by the vector boson masses.

To sum up, we have argued that, because of physics at the weak scale and
beyond, QED must be supplanted by new, ``nonrenormalizable" vertices such as
the one associated with $\alpha_{\mu F}$.  Thus, the actual theory that we are
dealing with takes the form
\beq
\eqalign{\leff=&\lqed+\lnr,\cr
\lnr\equiv&{1\over\Lambda}\sum\alpha_i^{(5)}\ocal_i^{(5)} +
{1\over{\Lambda^2}}\sum\alpha_i^{(6)}\ocal_i^{(6)} + . . . ..\cr}
\eeq
where the $\ocal_i^{(N)}$ represent local operators of dimension $N$,
corresponding to vertices represented by the coupling constants
$\alpha_i^{(N)}$.\footnote{\ninerm\baselineskip=11pt As if the new physics is
associated with a single common new energy scale, appropriate factors of
$\Lambda$ have been extracted so that each $\alpha_i^{(N)}$ are dimensionless.
Of course, there may be a variety of scales of new physics, so that the
inferred magnitudes or limits on the couplings must be interpreted in a
specific context.}

Having said that the true theory is not so simple as QED alone, we must ask in
what respect QED is self-consistent and how, for example, can it predict
$g_\mu-2$ when, in fact, there is an independent, elementary coupling constant
in $\lnr$ that is precisely of this form?  The answer is that QED, including
its loop corrections, represents physics at scales {\elevenit below} the scale
of new physics $\Lambda$.
\vskip0.2in

\setbox3=\vbox {\hsize=1.5truein
\epsfxsize=\hsize
\epsffile{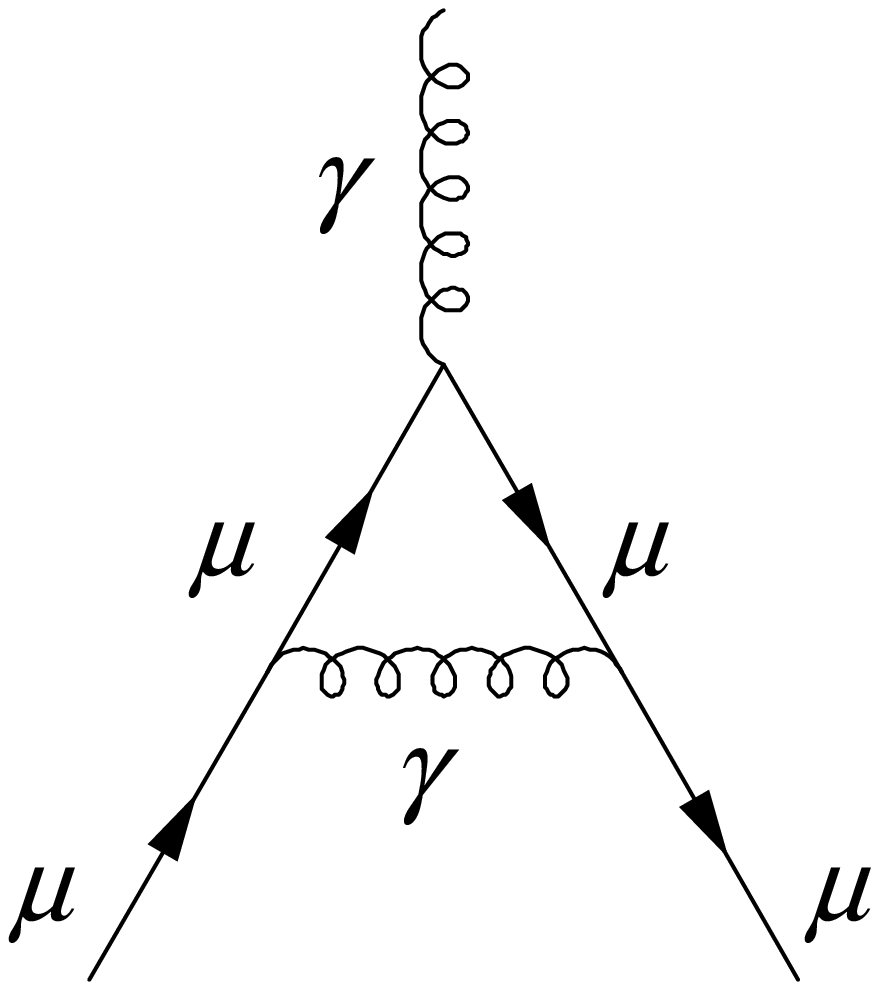}}

\vbox {\hbox to \hsize { \hfill \box3 \hfill }}

\centerline{\tenrm QED contribution to the anomalous magnetic
moment.}
\nobreak\centerline{\tenrm Fig.~2}
\vskip0.2in
\noindent Schwinger's famous prediction of $\alpha/2\pi$ for
the one-loop contribution (Fig.~2) is cutoff independent by virtue of the
renormalizability of QED.  At the same time, from our present perspective, it
should be regarded as uncertain by an amount of order $(m_\mu/\Lambda)^2$
since the physics at momentum scales above $\Lambda$ has not been correctly
represented by the renormalizable terms we associated with QED alone.  Stated
otherwise, one must add to the loop corrections the ``direct" contributions
associated with $\alpha_{\mu F}.$  It is only to the extent that Schwinger's
result is large compared to the size of this elementary vertex that QED
radiative corrections may be tested.

Given that there is an elementary Pauli interaction illustrated by Eq.~(6),
with vertex proportional to $\alpha_{\mu F},$ should it be included in
radiative corrections?  Being ``nonrenormalizable," in the traditional sense
of the term, the more times this vertex occurs, the greater the associated
degree of divergence of the graph.  However, familiar power-counting arguments
can be used to show that such divergences only contribute to local operators,
so that their contributions simply renormalize other parameters of the
effective field theory.  Thus, for example, the self-energy diagram of Fig.~3
will produce a correction to the muon mass which, like the bare mass, is
itself not directly observable.
\vskip0.2in

\setbox4=\vbox {\hsize=1.5truein
\epsfxsize=\hsize
\epsffile{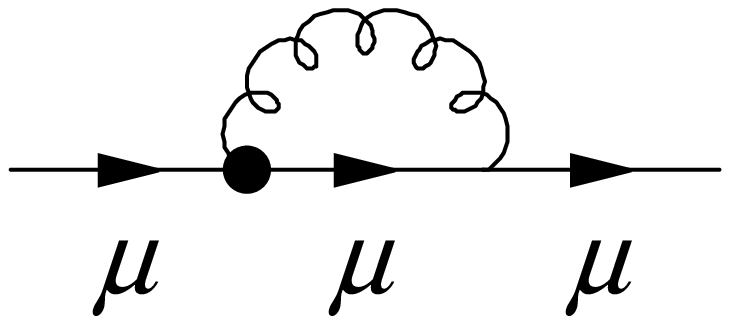}}

\vbox {\hbox to \hsize { \hfill \box4 \hfill }}

\centerline{\tenrm Anomalous vertex in a self-energy correction.}
\nobreak\centerline{\tenrm Fig.~3}
\vskip0.2in
\noindent Indeed, the most convenient method of renormalizing the full
effective Lagrangian Eq.~(8) is to assume that the renormalized parameters,
defined by minimal subtraction or another prescription, already include all
divergent contributions, not from loops containing just renormalizable
vertices but containing any vertices.  Thus, an effective field theory is
renormalizable, but an infinity of independent counterterms are required.  As
physical energy scales approach $\Lambda,$ increasingly many terms in $\lnr$
are required until the point is reached where the new physics can no longer be
accurately represented in terms of local operators of light fields.
Typically, new particles with masses of order $\Lambda$ are produced whose
propagation and interactions are not represented by $\leff.$

In the preceding, we spoke only of leptons, but, of course, a similar
calculation of the electromagnetic structure of the proton would break down at
a scale of $\Lambda_{QCD},$ even below the proton mass, revealing the
modifications due to strong interactions.  These too afflict leptonic
calculations, but, fortunately (for us and for Schwinger,) they begin to
contaminate QED predictions only at the two-loop level.  This illustrates that
the correlation between the sensitivity to new physics and its scale can be
subtle, and we must be careful in interpreting our results.

\vglue 0.6cm
{\elevenbf\noindent 3. The Standard Model and Beyond}
\vglue 0.4cm
\setcounter{footnote}{1}

Now let us focus on our main topic, the physics beyond the SM and its
implications for vector boson couplings.  Most of the lessons have been
illustrated in the preceding discussion of QED, which is why we took such
pains to explore it.  Conceptually, we replace QED with the SM:
\beq
\eqalign{U_1^Q&\Longrightarrow\sutwoone,\cr
\lqed&\Longrightarrow\lsm\cr
\leff&=\lsm+\lnr.\cr}
\eeq
Now $\lnr$ represents the physics the lies beyond the SM.
The assumption here is that there are no ``light" particles below some scale
$\Lambda$ other than the ones we know and, possibly, the Higgs boson.   This
assumption could be wrong, e.g., there could be supersymmetric partners with
masses not very different from these.  In that case, one would replace $\lsm$
with the minimal SUSY model $\lcal_{MSSM}$ and begin again to seek the physics
beyond expressed by $\lnr.$  But we shall suppose that the SM takes the form
\beq
\eqalign{\lsm=&\lcal_g\quad+\quad\lcal_m\quad+\quad\lhiggs,\cr
\lcal_g=&-{1\over2}{\elevenbf Tr}W_\mn^2 -{1\over4}B_\mn^2 +
\lcal_{g.f.}+\lcal_{F-P},\cr
&B_\mn\equiv\partial_\mu B_\nu-\partial_\nu B_\mu,\cr
&W_\mn\equiv\partial_\mu W_\nu-\partial_\nu W_\mu+ig_2[W_\mu,\,W_\nu],\cr
&W_\mu\equiv W_\mu^a{{\tau^a}\over2},\>\>\tau^a={\rm Pauli\>\>matrices},\cr
\lcal_m=&\sum_{n} \overline{\psi_{nL}}i\Dslash{D}\psi_{nL} +
\overline{\psi_{nR}}i\Dslash{D}\psi_{nR}
+\sum_{m,n}(y_{mn}\overline{\psi_{mL}}\phi\psi_{nR}+c.c.),\cr
D_\mu\psi_{nL}\equiv&\bigl(\partial_\mu+ig_2 W_\mu+ig_1 B_\mu
Y_{nL}\big)\psi_{nL}\cr
D_\mu\psi_{nR}\equiv&\bigl(\partial_\mu+ig_1 B_\mu
Y_{nR}\big)\psi_{nR}\cr}
\eeq
The form of the Higgs Lagrangian $\lhiggs$ is unknown at present, but, in the
``minimal" SM, it consists of a single Higgs doublet $\phi$ together with its
requisite gauge interactions.  I have chosen to include the Yukawa
interactions of such a field, which give rise to the CKM mixing angles, in
our ``matter" Lagrangian $\lcal_m$, but that would be modified in
nonminimal models.  In fact, there is no compelling reason to restrict the
number of Higgs doublets to one; there could be one doublet per generation
or, as in supersymmetric models, two doublets, one giving mass to
$T_3=+1/2$ fermions and the other, to $T_3=-1/2$ fermions.  The Higgs could
also be a composite field resulting from new strong interactions, as in
technicolor models, in which case a nonlinear representation is often used,
as will be reviewed subsequently.  The immediate goal of electroweak
experimentation is to unravel the nature of the Higgs sector and to uncover
whatever can be deduced about the physics that lies beyond, embodied here
in $\lnr.$

As in QED, gauge invariance may not be {\elevenit explicitly} broken without
severe consequences for consistency of this whole framework.   As previously,
$\sutwoone$ gauge symmetry provides both for the correct number of degrees of
freedom carried by the gauge fields and for the universal character of the
gauge interaction.  Indeed, non-abelian universality is even stronger than the
abelian case.  While one may in principle vary the hypercharge coupling
strength $g_1$ for each field,\footnote{\ninerm\baselineskip=11pt These
abelian couplings are constrained by anomaly cancellation, but, given the
number of different fields, much arbitrariness remains.} the nonlinear
character of the non-Abelian symmetry implies that there can be but a single
non-Abelian coupling strength $g_2$ relating cubic and quartic self-couplings
of the $W_\mu$ field as well as its coupling to fermions and scalars.  Gauge
invariance insures that this universality among Lagrangian parameters will be
manifested by the physical couplings as well.  It must be remembered that all
these consequences are nontrivial, to say the least, and must be kept in mind
if one begins to tinker with the dynamics of gauge fields.  This is why we do
not give masses to the weak vector bosons simply by adding mass terms
\beq
M_W^2\,{\elevenbf Tr}\,W_\mu^2+M_B^2\, B_\mu^2
\eeq
The {\elevenit raison d'etre} for $\lhiggs$ is to give masses in a gauge
invariant manner, via spontaneous symmetry breaking and the Higgs mechanism.
But there are two approaches to the description of $\lhiggs$:\nextline (1)~The
conventional approach involving one (or more) relatively light Higgs doublet
($m_H\lsim800~GeV$.) In the minimal model with a single doublet,
\beq
\eqalign{\lhiggs=&(D_\mu\phi)^\dagger D^\mu\phi + m^2\phi^\dagger\phi
-{\lambda\over4}(\phi^\dagger\phi)^2,\cr
D_\mu\phi=&\bigl(\partial_\mu+
ig_2 W_\mu-i{{g_1}\over2}B_\mu\big)\phi\cr}
\eeq
(2)~The second approach, although it arose historically by analogy with chiral
perturbation theory for pions,\cite{georgi} may also be thought of as a way of
generalizing St\"uckelberg's trick to non-Abelian theories.  The essence of
his method is to replace the gauge fields by their gauge transforms,
introducing in the process auxiliary scalar fields, now called would-be
Goldstone bosons, as dynamical fields that are eaten to form the longitudinal
components of the massive vector particles.  That can be done via a nonlinear
realization of the symmetry, introducing a $2\times2$ matrix $\Sigma$
according to
\beq
\eqalign{\Sigma\equiv&
\exp{ ( { { i\vec{\tau}\cdot\vec{w} } / {v} } ) },\cr
D_\mu\Sigma=&\partial_\mu\Sigma+i{{g_2}}W_\mu\Sigma-
i{{g_1}\over2}\Sigma B_\mu\tau_3\cr}
\eeq
The $\vec{w}$ is composed of the three would-be Goldstone fields, and
$v\approx 246~GeV$ is the electroweak energy scale.  By construction, $\Sigma$
transforms linearly under $\sutwoone$
\beq
\Sigma\to e^{ ( {
{-i\vec{\theta}_L\cdot\vec{\tau} } / {2} } ) }\,\Sigma\, e^{ ( { {
i{\theta_R}\tau_3}  / {2} } ) }\equiv
U_L(\theta_L)\>\Sigma\>U_1^Y(\theta_R)^\dagger
\eeq
This approach includes the SM on scales below $m_H.$
The corresponding matrix of Goldstone fields can be obtained from the
SM Higgs field $\phi$ by the association
\beq
\Sigma\equiv {{\sqrt2}\over v}\pmatrix{\widetilde{\phi}\>\> \phi}
\eeq
where $\widetilde{\phi}\equiv i\tau_2\phi^*.$  Accordingly, this description in
terms of the Goldstone bosons alone is more appropriate for the case of a very
heavy or ``strongly interacting" Higgs boson.

Defining $V_\mu\equiv\Sigma^\dagger (D_\mu\Sigma),$
the vector boson masses can be introduced in analogy with St\"uckelberg as
\beq
\lcal_2\equiv {{v^2}\over 4} {\rm Tr}\bigl(V_\mu V^\mu\bigr) +
\beta_1 {{v^2}\over8}\bigl[{\rm Tr}(\tau_3V_\mu)\bigr]^2.
\eeq
To see that these are simply gauge-invariant vector masses, go to the
``unitary gauge" where $\Sigma=1,$ in which case,
\beq
\lcal_2={{g_2^2v^2}\over4}M_W^2 \wcal_\mu^\dagger \wcal^\mu +
{{M_W^2}\over{2\cos^2\theta_w}}(1-\beta_1)Z_\mu^2,
\eeq
where
$\wcal_\mu\equiv{1\over{\sqrt2}}(W_\mu^1-iW_\mu^2)$ is the $W^-$ field;
$Z_\mu=\cos\theta_w W_\mu^3-\sin\theta_w B_\mu$, the $Z^0$ field, $M_W\equiv
g_2v/2,$ and $\theta_w$, the usual weak mixing angle defined by
$\tan\theta_w\equiv g_1/g_2.$  In tree approximation, obviously $M_W$ is the
$W^\pm$ mass, and its relation to $M_Z$ is parameterized in terms of a
parameter usually called $\rho$:
\beq
\rho\equiv({{M_W}\over{M_Z\cos\theta_w}})^2.
\eeq
Apparently, in tree approximation, we have $\rho=1/(1-\beta_1).$

The preceding two descriptions are the only ways known to describe massive
vector bosons in a gauge invariant manner.
Experimentally, $\rho-1$ agrees with SM predictions, including one loop
corrections, to better than 0.5\%,\cite{langacker} so it must be that
$\beta_1$ is small, $\beta_1\lsim1\%.$  This would seem unnatural unless
$\beta_1=0$ is associated with a higher symmetry.  In fact, the minimal SM
Lagrangian $\lsm$ does possess an $SU_2^R$ global symmetry in the limit that
hypercharge vanishes ($g_1=0$) and mass splittings within fermion doublets
vanish.  In the present context, that symmetry corresponds to embedding the
$U_1^Y$ hypercharge transformation on $\Sigma$ in an $SU_2^R$ group under
which
\beq
\Sigma\to \,\Sigma\, e^{ ( {{i\vec{\theta}_R\cdot\vec{\tau} } / {2} } ) }.
\eeq
In this limit, even after spontaneous symmetry breaking, there remains a
vector-like isospin symmetry that preserves $M_W=M_Z$.  Then, if the only
explicit breakings are via hypercharge and Yukawa couplings, one can show that
$\rho=1$ plus loop corrections.
\footnote{\ninerm\baselineskip=11pt  Nevertheless, these corrections are
proportional to the square of quark mass splittings, requiring $m_t\lsim
180~GeV.$  This custodial symmetry is not necessarily present when there are
more than one Higgs doublet.  It is, however, a natural consequence of chiral
symmetry in technicolor models.}

Possible physics ``beyond the SM" is represented by $\lnr.$  As with
$\lhiggs,$ depending on which of the two frameworks one adopts, it takes on
rather different forms with some different consequences.  It is a bit easier
to follow through the discussion in the linear case first before discussing the
nonlinear representation, to which we shall return.  In the case of a
relatively light Higgs with a linearly realized representation of the
symmetry, $\lnr$ consists of writing down all higher dimensional,
gauge-invariant operators involving the fields of the SM, as in Eq.~(8).
Assuming lepton and baryon conservation, there are no dimension~5 operators,
so $\lnr$ begins with dimension 6.  The generic scale of new physics $\Lambda$
is unknown, but is certainly larger than the weak scale $v\approx 246~GeV$ if
this linear representation is to make sense.  These have been catalogued
several years ago;\cite{buchwyl,others} even with a single fermion generation,
there are more than 80 independent dimension-6 operators.  So you might think
this parameterization would be overly general, but in fact, only a few affect
any given process and one may classify these operators to some extent by the
type of underlying physics that gives rise to them.

Since our interest here is in gauge boson structure, let us focus on those
dimension-6 operators in $\lnr$ that involve the vector fields and the Higgs
field only.\footnote{\ninerm\baselineskip=11pt I'll return shortly to the
non-linear realization.}  To simplify the discussion further, I will restrict
my attention to the CP-conserving operators.  Some of these operators
correspond to vertices contributing directly to the vector boson vacuum
polarization tensor.  These are the ones that have been probed experimentally
to some degree, so we'll begin with
them.\cite{buchwyl,others,derujula,gouren,zeppen}
\beq
\eqalign{
\ocal_\phi &\equiv |\phi^\dagger D_\mu\phi|^2,\cr
\ocal_{BW} &\equiv g_1g_2(\phi^\dagger  W_{ \mu\nu } \phi)  B^{ \mu\nu },\cr
\ocal_{DB} &\equiv {{g_1^2}\over2} \partial_\mu B_{\rho\sigma}
\partial^\mu B^{\rho\sigma}\cr
\ocal_{DW} &\equiv g_2^2 {\rm Tr}(D_\mu W_{\rho\sigma}D^\mu W^{\rho\sigma})
\cr}
\eeq
Since these modify the Born amplitudes for $e^-e^+$ annihilation to fermion
pairs at the $Z^0$, there already exist constraints on the magnitude of these
operators\cite{derujula,zeppen} as a result of the SLC/LEP-1 experiments.
These are most easily quoted as if only one of these operators were present,
and of course multiparameter fits will be weaker.  Assuming that, and
assuming that the scale of new physics is as low as possible for this
expansion to make sense, $\Lambda=v$, then the typical constraint on the
associated coupling constants $\alpha_i$ are
\begin{eqnarray}
 |\alpha_\phi |\ \lsim\ 0.3\% &\phantom{xxx}& |\alpha_{DW}|\ \lsim\ 3\%\\
 |\alpha_{BW}|\ \lsim\ 2\% &\phantom{xxx}& |\alpha_{DB}|\ \lsim\
 10\%\nonumber
\end{eqnarray}
As the operators in the first column explicitly break the custodial $SU_2^R$
symmetry, we would expect them to be small anyway.  While these are useful
numbers, at this level of accuracy, they give us no further insight into the
physics beyond the SM.\footnote{\ninerm\baselineskip=11pt One might think
that, if the Higgs sector were strongly interacting, these effects might be
enhanced, but, as we shall discuss later in this lecture, that is not likely
to be the case, with technicolor models also suggesting a magnitude of no more
than a few tenths of one percent.  Rather than explore that branch of
development immediately, it is pedagogically preferable to carry through
within the context of the linear case.}

There are other operators which are unobservable until Higgs boson
interactions are observed.  Such operators have been appropriately called
``innocuous"\cite{derujula} and include
\beq
\eqalign{ \ocal_{\widetilde{\phi\phi}}
&\equiv (\phi^\dagger\phi) (D_\mu\phi^\dagger D^\mu\phi),\cr
\ocal_{BB}&\equiv {{g_1^2}\over4}(\phi^\dagger\phi) B_{\mu\nu}B^{\mu\nu}\cr
\ocal_{WW}&\equiv {{g_2^2}\over4}(\phi^\dagger\phi)
{\rm Tr}(W_{\mu\nu}W^{\mu\nu})\cr}
\eeq
The point about these operators is that, if one goes to the unitary gauge and
simply replaces $\phi$ by its vacuum
expectation value, thereby suppressing the interactions of Higgs bosons,
these operators simply become wave function renormalizations and are,
therefore, unobservable.

A third class of operators are those whose vertices involve three or more
vector bosons and, therefore, will contribute to vacuum polarization only
through loops.  These include
\beq
\eqalign{
\ocal_{WWW} &\equiv g_2^3{\rm Tr}(W_{\mu\nu}W^{\nu\lambda}W_\lambda^\mu)\cr
\ocal_{B\phi} &\equiv i {{g_1}\over2}(D_\mu\phi)^\dagger B^{\mu\nu}
D_\nu\phi\cr
\ocal_{W\phi} &\equiv g_2(D_\mu\phi)^\dagger W^{\mu\nu} D_\nu\phi\cr}
\eeq
Contributions of these operators to tree diagrams have not been experimentally
tested, and, as we shall have more to say later, there are as yet no useful
constraints from considering them in loops either.
\vglue 0.2cm
{\elevenbf\noindent 4. Standard Parameterization of Deviations}
\vglue 0.2cm
\setcounter{footnote}{1}

Of primary interest for LEP-2 and the NLC is the nature of the triple-vector
boson couplings.  There are a variety of fields and tensor structures that
make this discussion unavoidably tedious.  In a now-standard
reference,\cite{hagiwa} the most general trilinear vector couplings were
displayed, assuming only Lorentz invariance.  We will restrict our attention
here to the CP-invariant terms only:
\beq
\eqalign{ &{\cal L}_{WWV}/g_{WWV}=
\>ig_1^V (\wcal_{[\mn]}^\dagger \wcal^\mu \vcal^\nu-h.c.) +i\kappa_V
\wcal_\mu^\dagger
\wcal_\nu \vcal^\mn\cr &+i{{\lambda_V}\over{M_W^2}}
\wcal_{[\lambda\mu]}^\dagger
\wcal^{[\mn]} \vcal_\nu^\lambda +g_5^V\epsilon^{\mu\nu\rho\sigma}
(\wcal_\mu^\dagger \partial_\rho\wcal_\nu-
\partial_\rho\wcal_\mu^\dagger \wcal_\nu) \vcal_\sigma,\cr }
\eeq
where $\wcal_\mu$ is the $W^-$ field,
$\vcal_\mu$ represents either the photon $\vcal=\gamma$ or the $Z^0$-boson
$\vcal=Z,$
$\vcal_\mn\equiv\partial_\mu \vcal_\nu-\partial_\nu \vcal_\mu$ and
$\wcal_{[\mn]}\equiv\partial_\mu \wcal_\nu-\partial_\nu \wcal_\mu$  are the
``Abelian field strengths," and the normalizations have been chosen to be
$g_{WW\gamma}\equiv e$; $g_{WWZ}\equiv e\cot\theta_w$.
\footnote{\ninerm\baselineskip=11pt   The last term involving $g_5^V$ is
separately C and P violating, and for some reason, is often dropped from these
discussions despite the fact that the electroweak theory violates these
symmetries maximally.  The vertices in ${\cal L}_{WWV}$ were assumed to be for
on-mass-shell vector bosons, which is sufficient for tree-level applications
for which this formalism was originally developed.}  Requiring electromagnetic
gauge invariance leads to $g_1^\gamma\equiv1,\quad g_5^\gamma=0$, and these
are generally assumed true.  The SM values for the rest are
\beq
g_1^Z=\kappa_\gamma=\kappa_Z=1,\quad \lambda_\gamma=\lambda_Z=g_5^Z=0.
\eeq

What is the relation between the previous effective Lagrangian formalism to
these conventional parameters?  By going
to unitary gauge in Eqs.~(20-23), one can extract a variety of
results, such as:\cite{gouren,zeppen}
\beq
g_5^Z=0,\qquad \lambda_\gamma=\lambda_Z.
\eeq
These results are specifically the result of the truncation of $\lnr$ to
dimension-6 operators and will be modified by higher dimensional operators.
If  experimental accuracy were sufficient to test these relations (about
which we are rather skeptical,) they would provide an indication of the
scale of new physics.  The actual relation for the so-called ``quadrupole
moment" is
\beq
{{\lambda_\gamma}\over{M_W^2}}={3\over2}{{g_2^2}\over{\Lambda^2}}\alpha_{WWW}
\eeq
This formula indicates that defining the dimensionless parameter $\lambda$ by
scaling out the mass $M_W$ can be quite misleading, since the natural scale is
set by $\Lambda$, the scale of new physics.\footnote{\ninerm\baselineskip=11pt
In this respect, it is analogous to the traditional definition of $g_\mu-2$ by
scaling out the muon mass, which is very misleading when it comes to
discussing the magnitude of contributions from new physics at high scales.}
The relations for the other vertices are rather more complicated, involving
many of the dimension-six operators.  However, the degree to which SLC/LEP-1
constrain those operators that contribute to the vacuum polarization tensor is
so great as to render them beyond reach for LEP-2 experiments.\cite{derujula}
So if one simply assumes that the operators that have not been constrained by
existing experiments may be much larger than those that are,\cite{zeppen} then
one finds that the three deviations $\Delta\kappa_\gamma,$ $
\Delta\kappa_Z,$ and $\Delta g_1^Z$ from their SM values may be expressed in
terms of the two couplings $\alpha_{B\phi}$ and  $\alpha_{W\phi}.$  This leads
to the non-trivial relation\cite{zeppen}
\beq
\Delta g_1^Z=\Delta\kappa_Z+\tan^2\theta_w\,\Delta\kappa_\gamma.
\eeq
This relation also depends on the additional assumption that dimension-8
operators are negligible compared to the dimension-6 ones, which, as we shall
see, is equivalent to assuming that the Higgs sector $\lhiggs$ is not strongly
interacting and can be represented linearly as in Eq.~(12).

As a matter of fact, I know of no model that satisfies this assumption that
the unconstrained dimension-6 operators are larger than those that have been
constrained by existing experiments.  Stated otherwise, I know of no model in
which the underlying new physics contributes at one-loop order, for example,
to the triple-vector boson vertices  but does not also contribute at least as
large a contribution to the vacuum polarization tensor.  The assumption
underlying Eq.~(28) is also implausible since not all the dimension-6
operators that I have written down are linearly independent, that is, the
choice of form of the higher order terms is somewhat arbitrary, because one
may replace some operators by others without changing any physical
consequences.\footnote{\ninerm\baselineskip=11pt This has been discussed at
some length, for example, in Ref.~\cite{derujula}.  For further discussion
of the theoretical underpinnings, showing that this arbitrariness in the
choice of operators may also be extended to their use in loop calculations,
see Ref.~\cite{arzt}.}  To illustrate how subtle this can be, a common
basis set of dimension-6 operators is one in which the operators on which
the relation Eq.~(28) are based, viz.,  $\ocal_{B\phi}$ and
$\ocal_{W\phi}$, do not occur!\cite{buchwyl}  So, while phenomenologically
permitted at present, I would not place much stock in this possibility.
This is a special case of the assumption in Ref.~\cite{derujula} that the
new physics does not point in ``blind directions," a terminology that I
haven't time to fully explain here.  \footnote{\ninerm\baselineskip=11pt
Loosely speaking, it is the assumption that new physics does not prefer one
basis set over another.  However, the absence of ``blind" directions is not
a general principle about operators arising from new physics, since it is
{\nineit not} the case that all models necessarily yield all interrelated
dimension-6 operators of comparable magnitude.  For example, depending on
the underlying theory, four-fermion interactions may either be suppressed
or enhanced relative to operators involving gauge bosons.}

\vglue 0.2cm
{\elevenbf\noindent 5.  Strongly Interacting Higgs}
\vglue 0.2cm
\setcounter{footnote}{1}

We have indicated previously that, at scales well below the Higgs mass, or
in models of dynamical symmetry breaking in which there may be no Higgs
boson at all, in which case a nonlinear representation of $\lhiggs$ is more
appropriate than the linear prescription usually employed in the SM.  Let
us now turn to the question of how, in the nonlinear framework, the physics
beyond the SM in $\lnr$ is to be represented.  Dimensionality of the field
no longer plays the key role that it did in the linear framework.  However,
Goldstone bosons are derivatively coupled, so their interactions are
proportional to their momenta.  Thus, one can perform a momentum expansion
whose scale $\Lambda$ is set by the scale at which their interactions
become strong.  This has formed the basis for chiral perturbation theory
for pions,\cite{weinberg,gassleut} where the expansion scale is set by the
lowest-lying resonances, the vector mesons.  On theoretical grounds, one
may argue that, in general, $\Lambda$ is not expected to be {\elevenit
larger} than $4\pi v$, where $v$ is the scale of symmetry
breaking.\footnote{\ninerm\baselineskip=11pt In the case of pions, the
corresponding scale $v\equiv f_\pi\approx 95 MeV$ whereas
$m_\rho=770~MeV.$} In this expansion, the leading terms, the mass terms of
Eq.~(17), were of ``chiral-dimension" 2.  In order to maintain manifest
$\sutwoone$ gauge-invariance, ordinary derivatives have been replaced by
covariant derivatives so that the gauge field is counted as dimension one,
as usual. There we inferred that, in such a theory, the mass terms that
explicitly break the custodial $SU_2^R$ symmetry must necessarily be small,
about 1\% of those that do not.  So we will expect that to be a
characteristic of this description, in other words, for the operators that
break the custodial symmetry, one may wish to associate a somewhat larger
scale than $4\pi v$ or to regard the natural size of their couplings to be
smaller than for the operators conserving $SU_2^R.$  In any case, instead
of $\lnr$, we would similarly add
\beq
\lcal_{new}={1\over{16\pi^2}}\sum \alpha_k\ocal_k,
\eeq
where we have extracted a conventional\cite{georgi} factor of $1/16\pi^2$ in
the definition of the coupling constants, but, as indicated, it should really
be a factor of $v^2/\Lambda^2$ where $\Lambda$ now represents the scale at
which the longitudinal vector bosons become strongly interacting.  The next
terms in this expansion are chiral-dimension 4, having two more derivatives:
it is a bit complicated to display a complete set of operators that are
linearly independent under application of the classical equations of motion.
Restricting ourselves to CP-conserving operators, ten were listed by
Longhitano\cite{long} a long time ago, and an eleven was recently pointed out
in a recent paper by Appelquist and
Wu.\cite{appelwu}\footnote{\ninerm\baselineskip=11pt Longhitano's catalogue
included 3 CP-violating operators, but Ref.~\cite{appelwu} lists 5 additional
ones.}  Because of space limitations, I cannot share all of them with you,
here, but three of the most important ones that, in the limit of zero
hypercharge, conserve custodial symmetry, and can therefore expected to have
the largest coefficients are, in the nomenclature of Ref.~\cite{appelwu},
\beq
\eqalign{ {\ocal}_3 & \equiv i g  Tr(W_{\mu\nu}[V^{\mu}, V^{\nu}])\cr
{\ocal}_4 & \equiv  [Tr(V_{\mu}V_{\nu})]^2\cr
{\ocal}_5 & \equiv  \alpha_5 [Tr(V_{\mu} V^{\mu})]^2,\cr}
\eeq
where $V_\mu$ was defined above Eq.~(16).

Whereas $\ocal_3$ (and 5 other unspecified $SU_2^R$-breaking operators)
contribute to a triple-vector boson vertex, the other two (plus 4 others that
break the custodial symmetry) give quartic gauge-boson vertices and higher.
These are the ones that involve interactions among the longitudinal vector
bosons themselves and interact like the Goldstone
bosons\footnote{\ninerm\baselineskip=11pt They survive even when you switch
off the gauge fields.}  so these are the ones that may be expected to be
strongest.  They are obviously important for discussing potentially strong
$WW$-scattering processes, for example, at the NLC or SSC, or for linear
colliders at extremely high energies,\footnote{\ninerm\baselineskip=11pt See
the session at this conference on the strongly interacting Higgs sector for
the phenomenological implications of these terms.}  but it is $\ocal_3$ that
is most relevant to modifications of vector boson production at NLC.  Now, you
may note that, if the matrix $\Sigma$ were replaced by the doublet Higgs field
$\phi$, $\ocal_3$ would be of dimension 8.  This illustrates a primary
phenomenological distinction from the linear representation; {\elevenit
operators that were of dimension-8 may, because of strong Higgs
self-interactions, become as important as operators of dimension-6.} Going to
the unitary gauge, it is straightforward (but tedious) algebra to obtain the
relations between these operators and the standard parameters.\cite{appelwu}.

\vglue 0.2cm
{\elevenbf\noindent 6.  Orders of Magnitude}
\vglue 0.2cm
\setcounter{footnote}{1}

As indicated earlier, the experimental situation and potential will be
summarized by Dr.~Miyamoto in the next lecture.  However, the sensitivity that
seems to emerge for  prospective facilities is
\beq
\eqalign{
{\tenrm LEP(200):} &|\kappa-1|, |\lambda|\> \lsim  10\%,\cr
{\tenrm NLC(500):} &|\kappa-1|, |\lambda|\> \lsim 1\%,\cr
{\tenrm NLC(1000):}  &|\kappa-1|, |\lambda| \lsim 0.1\%.\cr}
\eeq
Are there general theoretical or phenomenological arguments concerning the
magnitude of the deviations to be expected from the SM predictions?  Clearly,
each gauge field brings in a power of a gauge coupling, but, more importantly,
the triple vector boson operators are at least of one-loop order in any
underlying theory conserving $\sutwoone.$\cite{forth}  (This is not true for
the four-point coupling, for example.)  Therefore, it is natural to expect all
the $\alpha_i$ that contribute directly to triple-vector-boson vertices to
involve powers of the gauge couplings times at least a factor of ${1/16\pi^2}$
from the loop phase space factor.  Thus, for example, we expect
\beq
\alpha_{WWW} = { g^3 \over 16 \pi^2 }x_W , \quad
\alpha_{WB} = { g g' \over 16 \pi^2 }x_{WB},
\eeq
with the $x_i$ of order one.  The next question is how small could $\Lambda$
possibly be?
Certainly, in the usual, linear realization
of $\lhiggs$, the scale of new physics $\Lambda\sim v.$
These observations imply that the natural values to be expected for deviations
from the SM are no larger than a few tenths of one percent per loop:
\beq
\eqalign{|\kappa-1|\lsim\>&3\times 10^{-3}.\cr
|\lambda|\lsim\>&2\times 10^{-3}.\cr}
\eeq
Now it is reasonable to suppose that new physics involves several particles
that might contribute constructively or that $\Lambda$ may be a bit smaller
than $v$ so that the limits may be closer to 1\%.  Referring to the
anticipated experimental accuracy, Eq.~(31), we see that NLC(500) may just
begin to provide interesting and meaningful constraints.  LEP-2, on the
other hand, would at best rule out rather bizarre models that have hundreds
of particles adding up coherently.

Now, you might think that one could generate larger effects in models in which
there is a strongly interacting Higgs sector.  But, in fact, their Goldstone
character of the longitudinal modes means that, like pions, they couple
weakly, until a scale considerably larger than the vector mass.  Further,
unitarity serves to limit their strength, so the effective scale parameter
$\Lambda$ remains larger than the weak scale.  In the final
analysis, results on the overall strength are not expected to be very
different from the linear case.\cite{hiro}  This conclusion is supported by
the recent general analysis of Ref.~\cite{appelwu}.

\vglue 0.2cm
{\elevenbf\noindent 7.  Gauge Invariance; Conclusions}
\vglue 0.2cm
\setcounter{footnote}{1}

In this lecture, I have emphasized that $\sutwoone$ gauge symmetry strongly
constrains the possibilities for new physics.  This is actually a slightly
controversial assertion that I will now address.  A number of
people\cite{hiro,blone,knetter}  have remarked that the construction of the
chiral Lagrangian involving the non-linear representation of the underlying
symmetry  be regarded as a non-Abelian generalization of the St\"uckelberg
trick.  Thus, any non-gauge invariant Lagrangian can be willy-nilly assumed
to be the unitary gauge expression of a gauge invariant theory.  Therefore,
some have emphasized (especially Ref.~\cite{blone}) that gauge invariance
is essentially without content.  I will argue to the contrary.  Is there
really no difference between a weak vector boson and any other massive
vector boson, like the $\rho$-meson, the deuteron\cite{brodsky}, the
$J/\psi,$ etc.?\footnote{\ninerm\baselineskip=11pt Ref.~\cite{brodsky} is
extremely interesting in this context.  However, that derivation of the
``universal" character of the magnetic and quadrupole moments of a vector
particle depends on assumptions about the number of subtractions required
for convergence of their dispersion relations.  This will ultimately depend
on the short-distance structure of the theory, so I regard their conclusion
as a consistency condition rather than a proof of universality.}  This is a
central issue--the scale of the structure of the particle.  The essential
difference is that a gauge particle is a vector particle that can be
regarded as an elementary excitation of its field {\elevenbf over a range
of momentum that is large compared to its mass.}  Just because you can
write an arbitrary vector interaction in gauge invariant form does not mean
that is a useful description of nature beyond tree approximation or at
large momentum scales.  The only way that we know how to make a vector
particle whose mass is small compared to the scale of its structure is via
the Higgs mechanism.\footnote{\ninerm\baselineskip=11pt This is almost a
theorem.\cite{veltman}}  In the linear representation, it is manifest  that
the scale of new physics $\Lambda \gg v.$  In fact,  with an elementary
scalar field, there are fine-tuning or naturalness issues\cite{thooft} that
suggest that there is a scale above which this description will break down.
That scale cannot be larger than about $4\pi v,$ not so very different from
scale of compositeness in technicolor
models.\footnote{\ninerm\baselineskip=11pt  This is a different issue than
the so-called triviality bound on the Higgs mass that shows that the local
field theory becomes inconsistent for a Higgs mass larger than about
800~\gev.  For a review, see Chpt.~9 of Ref.~\cite{higgsbook}.}  But the
nonlinear realization is also only useful as an effective field theory if
successive terms of higher chiral dimension are suppressed over some energy
range large compared to the vector masses.  This is a primary reason we are
all so eager to probe the TeV region experimentally; the SM will change, at
least through the addition of superpartners if not even more dramatically.

If the weak bosons were not gauge particles in the sense that I have
described, then why are their couplings to fermions
universal?\footnote{\ninerm\baselineskip=11pt Universality for quarks is
complicated by the CKM mixing angles (and the fact that quarks cannot be
observed directly,) but the entire framework of phenomenology would be
modified dramatically by strong interactions were it not for the underlying
gauge symmetry.}  Radiative corrections calculated within the SM involve loop
integrals over a range of momenta up to the scale $\Lambda$.  The particles in
the loops are treated as structureless, point particles below that scale.
Moreover, the dependences on this scale for different loops are constrained by
gauge invariance.  Outside of this context, the SM would not be a good first
approximation insensitive to new physics at higher scales.

As SUSY models illustrate, this does not necessarily imply that all new
particles lie at masses much larger than $M_W$ and $M_Z$, but it also does not
mean that gauge invariance is without content.  In particular, the structure
of the gauge bosons are extremely sensitive to tampering.  There have been
many papers written showing this.  For example, the electromagnetic
self-energy of the $W^\pm$ would behave as $\Lambda^4$ if gauge-invariance
were explicitly broken.\cite{suzuki}   Such dependences are not directly
observable, since they simply multiply local operators and therefore
renormalize Lagrangian parameters, but these contributions would change our
present thinking considerably, because a theory that was not gauge invariant
would involve many more coupling constants and mass parameters than are in the
SM.  For example, in that framework, there would be no reason for the precise
relation between $M_W$ and $M_Z$, which would become independent parameters.
The issue is one of naturalness; while fine-tuning is always possible, it is
more likely that relations have reasons.  To be a good first approximation at
the scale $M_Z$, as they seem to be, the resulting corrections, like
Schwinger's old QED correction to $g-2$, must be no smaller than the
modifications coming from new physics.  Thus, the scale $\Lambda$ of new
physics ought to be larger than the vector masses.

This embedding of a non-gauge invariant Lagrangian in an underlying gauge
theory, regarding it simply as the unitary gauge expression of a gauge
invariant theory, does not dictate the underlying gauge group.  We do not, for
example, have to regard the general form in Eq.~(24) as necessarily coming
from an $\sutwoone$ gauge invariant theory.  We could choose another
embedding having this as the unitary-gauge expression, e.g. $(\otimes U_1)^4$.
Why does everyone choose to embed this in $\sutwoone$?  The statement that the
underlying symmetry is an $\sutwoone$ gauge symmetry has content.

Now I am aware that there is another point of view on these matters, one that
has been extensively developed by members of the BMT collaboration.\cite{bmt}
In this, one focuses exclusively on the vector boson interactions and attempts
to build up relations among the observables $\kappa$ and $\lambda$ without
assumptions about the underlying gauge-symmetry.  This approach employs, for
example, global symmetries or restrictions on the maximum power in the energy
dependence of vector-boson scattering amplitudes in order to realize relations
normally resulting from the effective Lagrangian approach that I have
discussed.  I simply have not been able to understand this alternate approach
as a consistent field theory beyond tree level.  Gauge invariance is
pervasive, and one may not simply assume that part of the Lagrangian
explicitly violates it while the remainder respects it.  The rules for
calculating loop corrections become ambiguous, and the non-gauge invariance of
one sector quickly spills over into the full theory.  The challenge to
proponents of this non-gauge-invariant alternative is to reproduce all that we
know about electroweak interactions while modifying the interactions of the
gauge bosons.  I simply do not know how to do it other than as I have
described within the framework of a gauge-invariant effective field theory.

I have already explained how, in the context of QED, the new divergences
induced simply renormalize other couplings in the effective field theory.  So
while there had been some confusion in the past about how to use these
nonrenormalizable operators in loop calculations, I think it is fair to say
that the confusion has passed,\cite{hiro,bltwo,zeppen} and it is now well
understood that an effective Lagrangian is in fact renormalizable in the sense
that all divergences may be absorbed in renormalizations of the coefficients
of one of its operators.  So the degree of divergence of loop corrections
or sensitivity to the cutoff $\Lambda$ can only be used for naturalness
arguments and are not directly observable.

The gauge-invariant effective field theory approach that I have outlined in
this lecture is the only way known that preserves the successes expressed
by the SM Lagrangian while allowing for and including potential effects of
new physics in a self-consistent manner.  Unfortunately, nature has
conspired to make new physics difficult to find, given our limited accuracy
and energy, so that, if the threshold for new physics is at energies beyond
our means, it is very difficult to find evidence of it through virtual
effects.  This is why the SM has withstood all tests to date.  While the
interactions among gauge bosons are of great interest, I am afraid the
situation is unlikely to be different for the weak vector boson
interactions, especially the triple-vector-boson couplings that can be
probed at NLC.  Although NLC can place interesting constraints on gauge
boson interactions, given the limited accuracy that can be realistically
anticipated, our best hope would seem to lie not in virtual effects, but in
crossing the threshold for direct production of new particles.

\vglue 0.2cm
{\elevenbf\noindent 8.  Acknowledgements.}
\vglue 0.2cm
I would like to thank the organizers of this Workshop for the opportunity
to speak and to the local organizers for the flawless arrangements that
have made this Workshop such an enjoyable experience.  I have benefited
from discussions with so many people here that I hesitate to single anybody
out, but I would especially like to thank Drs.\ Barklow, Hagiwara,
Miyamoto, Schildknecht, and Zeppenfeld.  This work was supported in part by the
U.S. Department of Energy.


\vglue 0.2cm
{\elevenbf\noindent  References}
\vglue 0.2cm


\begin{thebibliography}{9}
\bibitem{hiro} M.B. Einhorn and J. Wudka in {\elevenit Proceedings of the
Workshop on Electroweak Symmetry Breaking,} Hiroshima, Nov.~12--15, 1991,
(World Scientific, Singapore, 1992); also see our talk at ``Yale Workshop on
Future Colliders," Oct. 2--3, 1992, UM-TH-92-25, (Oct.\ 1992), to appear in
the Proceedings.
\bibitem{stuck} E.C.G. St\"uckelberg, Helv.\ Phys.\ Acta {\elevenbf 11} (1938)
299.
\bibitem{kinoshita} T. Kinoshita in{\elevenit Quantum Electrodynamics} ed.
T. Kinoshita (World Scientific, Singapore, 1990.)
\bibitem{georgi} See, for example, H.\ Georgi, {\elevenit Weak Interactions
and Modern Particle Theory,}\/ (Ben\-ja\-min/\-Cum\-mings, New York, 1984.)
\bibitem{cgg} M. Chanowitz, M. Golden, and H. Georgi, {\elevenit Phys.\ Rev.}\
{\elevenbf D36} (1987) 1490.
\bibitem{knetter} C. Grosse-Knetter and R. K\"ogerler,
Bielefeld Preprint BI-TP 92/56 (Dec., 1992); C. Grosse-Knetter,
Bielefeld Preprint BI-TP 93/17 (1993).
\bibitem{langacker} For a review, see P. Langacker, Lectures at TASI-92,
Boulder, (June, 1992)  (World Scientific, Singapore, to appear.)
\bibitem{blone} C.P. Burgess and D. London, MCGILL-92-04, Mar. 1992.
\bibitem{buchwyl} W.\ Buchm\"uller and D.\ Wyler, {\elevenit Nucl.\ Phys.}
{\elevenbf B268} (1986) 621; some other terms can be found in W.\ Buchm\"uller,
B.\ Lampe, and N.\ Vlachos, {\elevenit Phys.\ Lett.} {\elevenbf197B} (1987)
379.
\bibitem{others}  C.N. Leung, S.T. Love, and S. Rao, {\elevenit
Zeit.\ Phys.} {\elevenbf C31} (1986) 433;
C.J.C. Burgess and H.J. Schnitzer, {\elevenit Nucl.\ Phys.} {\elevenbf B228}
(1983) 464.
\bibitem{derujula} A. deR\'ujula et.al., {\elevenit Nucl.\ Phys.} {\elevenbf
B384} (1992) 3.
\bibitem{gouren} G.J. Gounaris and F.M. Renard, Montpellier PM-92-31 (July,
1992.)
\bibitem{zeppen} K. Hagiwara, et.al. MAD/PH/737 (March, 1993).
\bibitem{hagiwa} K. Hagiwara, et.al., {\elevenit Nucl. Phys.} {\elevenbf B282}
(1987) 253.
\bibitem{arzt} C. Arzt, U. Michigan UM-TH-92-28.
\bibitem{weinberg}   S.  Weinberg, {\elevenit Physica} {\elevenbf A96} (1979)
327.
\bibitem{gassleut} J. Gasser and H.  Leutwyler, {\elevenit Ann.\ Phys.}
{\elevenbf 158} (1984) 142; {\elevenit Nucl.\ Phys.} {\elevenbf B250} (1985)
465; \ibid {\elevenbf B250} (1985) 517; \ibid {\elevenbf B250} (1985) 539.
\bibitem{long} A. Longhitano, {\elevenit Phys.\ Rev.} {\elevenbf D22} (1980)
1166; {\elevenit Nucl.\ Phys.} {\elevenbf B188} (1981) 118.
\bibitem{appelwu} T. Appelquist and G-H Wu, YCTP-P7-93 (April, 1993).
\bibitem{holdom} B. Holdom, {\elevenit Phys.\ Lett.} {\elevenbf B258} (1991)
156.
\bibitem{forth}C. Arzt, M. Einhorn, and J. Wudka, in preparation.
\bibitem{veltman}  M. Veltman, {\elevenit Acta.\ Phys.\ Polonica} {\elevenbf
12} (1981) 437.
\bibitem{thooft}G. 't~Hooft, Lecture III  in {\elevenit Recent Developments in
Gauge Theories,} Proc. NATO ASI, Carg\`ese, France, Aug. 26--Sept. 8 1979, G.
't~Hooft et.al. (eds.) (Plenum, N.Y., 1980).
\bibitem{higgsbook}  {\elevenit The Standard Model Higgs Boson,}
ed. M.B. Einhorn, (North-Holland, Amsterdam, 1991.)
\bibitem{aew} C. Arzt, M. Einhorn, and J. Wudka, U. Michigan UM-TH-92-17, Nov.
1992.
\bibitem{brodsky} S.J. Brodsky and J.R. Hiller,  {\elevenit Phys.\ Rev.}
{\elevenbf D46} (1992) 2141.
\bibitem{suzuki} M.\ Suzuki, {\elevenit Phys.\ Lett.} {\elevenbf B143} (1984)
237; see also A.\ Cohen, H.\ Georgi, and B.\ Grinstein, {\elevenit Nucl.\
Phys.} {\elevenbf B232} (1984) 61.
\bibitem{bmt}  Bielefeld-Montpellier-Thessanloniki Collaboration.  See G.
Gounaris et.al., in {\elevenit $e^+e^-$ Collisions at 500 GeV: The Physics
Potential,} ed. P.M. Zerwas, DESY 92-123B (August, 1992), p. 735ff.   See
also Dr.~Schildknecht's contribution to this Workshop and, e.g., the recent
papers {\elevenit Phys.\ Lett.} {\elevenbf B302} (1993) 309; ; M. Bilenky,
et.al., Bielefeld BI-TP 92/44 (February, 1993).
\bibitem{bltwo} C. P. Burgess and D. London, {\elevenit Phys.\ Rev.\
Lett.} {\elevenbf 69} (1992) 3428.
\end{thebibliography}
\end{document}